\documentstyle[prl,aps,multicol]{revtex}

\begin{document}
\draft

\title{Disordered Totally Asymmetric Simple Exclusion Process: 
Exact Results}

\author{
Kiran M. Kolwankar${^{1}}$\cite{eadd:kmk} and
Alexander Punnoose${^{2}}$\cite{eadd:ap} \\
${^{1}}$Department of Mathematics,
Indian Institute of Science,
Bangalore, 560 012, India\\
${^{2}}$Department of Physics,
Indian Institute of Science,
Bangalore, 560 012, India\\
}

\date{Version: July 15, 1998; printed \today}
\maketitle

\begin{abstract}

We study the effect of quenched spatial disorder on the
current-carrying steady states of the totally asymmetric
simple exclusion process with spatially disordered jump
rates. The exact analytical expressions for the steady-state
weights, and the current are found for this model in one
dimension. We demonstrate how these solutions can be
exploited to study analytically the {\em exact} symmetries
of the system.  In particular, we prove that the magnitude
of the  steady-state current is left invariant when the
direction of all the allowed particle jumps are reversed.
Or equivalently, we prove that for any given filling and
disorder configuration, particle-hole transformation is an
exact symmetry that leaves the steady-state current
invariant.  This non-trivial symmetry was recently
demonstrated in numerical simulations by Tripathy \& Burma
(preprint cond-mat/9711302).  

\end{abstract}

\begin{multicols}{2}

%
%

Driven diffusive systems have been a subject of extensive
studies in recent years~\cite{ref:schmittmann/zia:95}. A key
ingredient of these systems is the presence of a driving
field due to which the underlying dynamics do not
generically obey detailed balance, leading to steady states
with non-vanishing currents. In the absence of a unifying
theme that encompasses our understanding of nonequilibrium
phenomena, these systems  offer a relatively simple
framework within which systems far from thermal equilibrium
may be studied.

Driven diffusive systems further provide the possibility of
studying the intriguing interplay of disorder, interactions
and drive for a wide range of parameter values like the
degree of disorder, filling and drive
strength~\cite{ref:fisher:87}.  These expectations are not
unfounded:~(i)~Driven diffusive systems in the absence of
disorder have been studied extensively and have revealed
basic differences between equilibrium and nonequilibrium
systems~\cite{ref:schmittmann/zia:95,ref:spitzer:70,ref:liggett:85}.
For instance, it has been demonstrated that spontaneous
symmetry breaking
(SSB)~\cite{ref:evans/foster/godreche/mukamel:95,ref:godreche/etal:95}
and phase separation can occur in one dimensional asymmetric
exclusion processes~\cite{ref:lahiri/ramaswamy:97}. In
contrast it is well known that one dimensional systems in
thermal equilibrium with short range interactions do not
exhibit phenomena like SSB and phase
separation~\cite{ref:landau/lifshitz:80};~(ii)~Also, systems
with disorder and drive but no interactions between
particles are well studied and
understood~\cite{ref:bouchaud/georges:90}.

A number of physical situations involving flow in random
media require an understanding of disorder driven diffusive
systems of interacting
particles~\cite{ref:narayan/fisher:94}. Most of the
understanding gained in these system have been largely based
on numerical simulations. Analytical characterization in
terms of the exact steady state measures in systems {\em
without} translational invariance have been found only in
the case of the disordered drop-push
model~\cite{ref:tripathy/barma:97}.

An interesting class of models for which there exists {\em
no} analytical characterization of the steady-state weights
or the steady-state current is the disordered totally
asymmetric simple exclusion process (DTASEP). Even the
single disorder case has not been amenable to an analytic
treatment.  Extensive numerical and mean-field studies have
been done and a number of interesting features have been
highlighted
\cite{ref:tripathy/barma:97,ref:janowsky/lebowitz:92,ref:schutz:93}.

In this work we derive, for the first time, formally exact
solutions for the steady-state weights, and current for the
DTASEP model in one dimension. These solutions are valid for
arbitrary disorder and particle fillings.  We further show
how these solutions can be exploited to study the exact
symmetries of the model by studying the transformations that
leave the magnitude of the steady-state current invariant.
In particular we show that the model possesses an exact
particle-hole symmetry for any {\em given} disorder
realization; a result that was observed in numerical
simulations by Tripathy \& Barma
\cite{ref:tripathy/barma:cmt97}. In addition we prove the
``obvious'' result that the steady-state current is a
constant across each bond. Although this result is
intuitively obvious, given that the dynamics conserves the
number of particles, to the best of the authors knowledge no
explicit proof of this statement for a disordered lattice
model exists.

%
%

\medskip 

\noindent{\bf Definition of the model:} 
The DTASEP model is defined on a 1D lattice of length $L$
with periodic boundary conditions. Each site can hold either
1 or 0 particle. Each bond $(i,i+1)$ of the lattice is
assigned a quenched random rate $\alpha_i$ chosen
independently from some chosen probability distribution. The
evolution is governed by random sequential dynamics defined
as follows: in a time interval $dt$ the particle attempts to
hop, with probability $\alpha_idt$ to its neighboring site
$i+1$. We consider the case in which the jumps are allowed
only in one direction (to the right) and is the same for all
bonds. In addition, the move is completed if and only if
site $i+1$ is unoccupied. The time averaged steady-state
current $J_i$ in the bond $(i,i+1)$ is given by:

\begin{equation}
\label{eqn:ji}
J_i=\alpha_i\langle n_i(1-n_{i+1}) \rangle
\end{equation}

\noindent where $n_i$ counts the number of particles at site
$i$.

Since for every particle hopping to the right, a ``hole''
jumps to the left, implies interchanging all the particles
for holes (charge conjugation~-~C) and reversing the
direction of hopping (time reversal~-~T) leaves the steady
state current, up to a sign, the same. Hence the
steady-state current is symmetric under a CT transformation.
This symmetry is valid in general: in any dimension; in the
presence of disorder; and in the case when  the particle can
hop along any direction with finite probabilities (DASEP).

In Eq.~(\ref{eqn:ji}), if all the $\alpha_i$'s are put equal
to the same constant $\alpha$, then it follows that the
steady state current $J_0$, which is assumed to be the same
on each bond and hence $J_0=L^{-1}\sum_i J_i$, is invariant
under $n_i\rightarrow 1-n_i$ for all $i$. Hence for the
clean totally asymmetric simple exclusion process in 1D
(TASEP), charge conjugation (C) in itself leaves the
steady-state current invariant. Since CT symmetry is always
true, it follows that for the TASEP model in 1D both charge
conjugation and time reversal (T) symmetries individually
holds.

The surprising observation made in the numerical simulations
in Ref.~\cite{ref:tripathy/barma:cmt97} was that, charge
conjugation symmetry was found to hold even in the presence
of disorder in the DTASEP model in 1D. This symmetry as is
obvious from Eq.~(\ref{eqn:ji}) is not expected to hold when
all the $\alpha_i$'s are taken to be random. Convincing
numerical evidence has been offered, although a general
proof of the validity of this symmetry for all fillings and
disorder configurations has not been obtained.

We derive below for this model a formally exact expression
for the steady-state probability density and the current
$J_0$.  Using these expressions we show that the observed
reflection symmetry holds in general for all fillings and
disorder configuration which in turn implies the symmetry of
the steady-state current under charge conjugation.

%
%

For $L$ sites with $N$ particles, the number of
configurations $M=~^LC_N$. Here $^LC_N={L!/(L-N)!N!}$ is the
number of ways $N$ particles can be distributed amongst $L$
sites with a maximum of only one particle per site. The
dynamics of these configurations, for a given realization of
the quenched bond variables ${\bf
R}=\{\alpha_1,\alpha_2,\cdots,\alpha_L\}$, are given by the
rate equation:

\begin{equation} 
\label{eqn-transition} 
{dP_m\over dt} =
\sum_n T(n\rightarrow m)P_n - \sum_{n'}T(m\rightarrow n')P_m
\end{equation}

\noindent where $T(m\rightarrow n)$ gives the transition
probability from configuration $m \rightarrow n$.  This
expression can be conveniently expressed in matrix notation
as:

\begin{equation} 
\label{eqn-rate} 
{d {\bf P}(t)\over dt }= W {\bf P}(t)
\end{equation}

\noindent with $W_{mn}=T(n\rightarrow m) \mbox{ and }
W_{mm}=-\sum_{n'} T(m\rightarrow n')$.  Since ${\bf
1}\cdot{\bf P}(t)=\sum_m P_m(t) = 1 ~\forall~ t $
(normalization), it follows that ${\bf 1}\cdot W=0$.  This
implies that $W$ has a zero eigen-value with left
eigen-vector ${\bf 1}$. Hence there also exists a right
eigen-vector ${\bf P}$ such that $W{\bf P}=0$.  This defines
the steady-state solution of the problem for a given choice
of the disorder ${\bf R}$.

The steady-state of the problem considered above is
characterized by a uniform current $J_0$ across each bond.
Given the solution ${\bf P}$, the current across, say, site
$i$ and $i+1$, is given as $\alpha_i \sum P_m = J_0$, where
the sum is {\em only} over the set of configurations with a
particle on site $i$ and a hole on site $i+1$ (see
Eq.~\ref{eqn:ji}). 

The above sum for all sites $i$ can be combined and
conveniently written in matrix notation by defining a matrix
$G$ such that $G\cdot{\bf P}= J_0 ( 1/\alpha_1,
1/\alpha_2,\cdots,1/\alpha_L)^T\equiv J_0(1/\alpha)^T$.  The
entries of $G$ for a given row $i$ (corresponding to site
$i$) has 1 corresponding to the configurations with a
particle at site $i$ and a hole at site $i+1$ and 0
otherwise.  The order of the $G$ matrix is therefore
$(L\times M)$ since there are $L$ sites and $M$
configurations.  The $G$ matrix can be expanded to an
$(M\times M)$ matrix by augmenting an $(M-L) \times M$ zero
matrix to $G$, such that:

\begin{equation}
Q\cdot {\bf P}\equiv 
\left( \begin{array}{l} 
G_{\mbox{{\tiny L$\times$M}}} \\ 
0_{\mbox{{\tiny (M-L)$\times$M}}}
\end{array} \right)\cdot {\bf P} =
J_0 
\left( \begin{array}{c} 
1/\alpha \\ 0 
\end{array} \right) \equiv
J_0{\bf V}
\end{equation}

\noindent where for convenience we have defined new
variables $Q=(G,0)^T$ and ${\bf V}=(1/\alpha,0)^T$.  Hence
the three equations that determine the steady-state
distribution and current are:

\begin{mathletters} 
\begin{eqnarray} 
W\cdot {\bf P} &=& 0
\label{eqn-wp}\\ 
Q \cdot {\bf P}&=& J_0 {\bf V} 
\label{eqn-gp} \\
{\bf 1}\cdot {\bf P} &=& 1 
\label{eqn-norm}
\end{eqnarray} 
\label{eqn-wg} 
\end{mathletters}

If we assume that the $\mbox{row-rank}(W)=M-1$ and not
lesser, then there exists a one-parameter solution to
Eq.~(\ref{eqn-wp}). The most general form of the solution
will be of the form $\beta {\bf P}$, where $\beta$ is the
arbitrary parameter. The solution when substituted into
Eq.~(\ref{eqn-gp}) allows the $\beta$ factor to be absorbed
in the definition of $J_0$ by re-scaling $J_0\rightarrow
J_0/\beta$. Hence, the most general solution ${\bf P}\equiv
{\bf P}(J_0)$ will have  $J_0$ appearing as the only free
parameter. The value of $J_0$ is fixed by the normalization
condition given in Eq.~(\ref{eqn-norm}). Hence a unique
${\bf P}$ and $ J_0$ solves Eq.~(\ref{eqn-wg}).  

It should be noted that both Eq.~\ref{eqn-wp} and
Eq.~\ref{eqn-gp} cannot be inverted to obtain a solution for
${\bf P}$ and $J_0$ as neither $W$ nor $Q$ are invertible
matrices. Although, $W$ and $Q$ are independently
non-invertible, their sum $Q+W$ is invertible. The proof of
this is simple. Suppose $\mbox{row-rank}(Q+W)=r < M$. Then
the equation: 

\begin{equation} 
(Q+W)\cdot{\bf P}=J_0{\bf V} 
\label{eqn-wq}
\end{equation}

\noindent obtained by adding Eqs.~(\ref{eqn-wp}) and
(\ref{eqn-gp}) would have a general solution ${\bf
P}(J_0,\beta_1,\beta_2,\cdots,\beta_{M-r})$, i.e.,  in
addition to the free parameter $J_0$, there will be $M-r$
other free parameters. However, it was shown earlier that
the most general solution of ${\bf P}\equiv {\bf P}(J_0)$
can have only one free parameter $J_0$, that can be fixed by
normalization. It follows that $r=M$ which implies that
$(Q+W)$ is invertible.

Using Eq.~(\ref{eqn-wq}), we obtain ${\bf P}(J_0)=J_0
(Q+W)^{-1}\cdot{\bf V}$.  The steady-state current $J_0$ is
obtained using the normalization condition in
Eq.~(\ref{eqn-norm}) giving, $1/J_0={\bf 1}\cdot
(Q+W)^{-1}\cdot {\bf V}$.   Hence the final solutions are
given as:
\begin{mathletters} 
\begin{eqnarray} 
{\bf P}&=&{(Q+W)^{-1}\cdot {\bf V} \over {\bf 1}\cdot
(Q+W)^{-1}\cdot {\bf V}} \label{eqn-solp}\\ J_0&=&{1\over
{\bf 1}\cdot (Q+W)^{-1}\cdot {\bf V}} \label{eqn-solj}
\end{eqnarray} 
\label{eqn-sol} 
\end{mathletters}
\noindent Note that in general a linear combination: 

\begin{equation}
\label{eqn-eta}
(Q+\eta W)\cdot {\bf P}= J_0 {\bf V}
\end{equation}

\noindent could have been used. This is equivalent to the
equation $(Q+W)\cdot {\bf P} + (\eta-1)W\cdot {\bf
P}=J_0{\bf V}$.   Since ${\bf W}\cdot {\bf P}=0$ we get back
Eq.~(\ref{eqn-wq}) implying that ${\bf P}$ will be
independent of $\eta$ as it should be.

In arriving at the above solutions, we had assumed that in
steady-state, the current across each bond was the same
constant $J_0$. The arguments presented above clearly
indicate that there would be no unique solution had we
started by assuming that the currents were different. For
suppose the currents across the bonds were
$\{J_1,J_2,\cdots,J_L\}$, then the most general solution
would be of the form ${\bf P}(J_1,J_2,\cdots,J_L)$. Since
there are $L$ unknowns but only one normalization condition,
a unique solution would not exist.  We therefore conclude
that the currents in steady state are a constant across each
bond.

We now use these solutions to prove the reflection symmetry
of the current observed in
Ref.\cite{ref:tripathy/barma:cmt97}. It was, as stated
earlier, observed in Monte Carlo simulations that, for a
{\em given} realization of the quenched bond variables ${\bf
R}=\{\alpha_1,\alpha_2,\cdots,\alpha_L\}$, the steady
state-current was the same whether the particles were
allowed to jump to the right or to the left. 

We denote the set of quenched random bond variables when the
particles are allowed to jump to the left as ${\bf
\overline{R}}=\{\alpha_1,\alpha_2,\cdots,\alpha_L\}$. In
${\bf \overline{R}}$ the magnitude and the ordering of the
$\alpha_i$'s are the same as in ${\bf R}$ although the
direction of hopping has been reversed. Let ${\bf
\overline{P}}$ be the steady-state weights and
$\overline{W}$ and $\overline{Q}$ the corresponding
matrices such that $\overline{W}\cdot{\bf \overline{P}}=0$
and $\overline{Q}\cdot{\bf \overline{P}}=\overline{J}_0{\bf
V}$. The ${\bf V}$ vector in both cases are the same and
$\overline{J}_0$ is the appropriate current.

The question we ask here is: Does there exist an invertible
matrix $S$, such that:
\begin{mathletters} 
\label{eqn-qp} 
\begin{eqnarray} 
Q\cdot S &=&
\overline{Q} 
\label{eqn-qs}\\ 
S^{-1}\cdot {\bf P} &=&  
{\bf \overline{P}} 
\label{eqn-sp} 
\end{eqnarray} 
\end{mathletters}
\noindent The existence of such an $S$ would imply that
$\overline{J}_0=J_0$, since:

\begin{equation} 
\label{eqn-j0}
\overline{Q}\cdot {\bf \overline{P}} 
= Q\cdot SS^{-1}\cdot {\bf P}=
Q\cdot {\bf P} 
\end{equation}

We first derive the properties that $S$ should satisfy.
Since each row of $G$ contains as many 1's as the number of
configuration for which there exists a particle at site $i$
and a hole at site $i+1$, we have $G\cdot {\bf
1}=~^{L-2}C_{N-1}{\bf 1}$. The same is true of
$\overline{G}$, although the exact location of each of the
1's will differ since the direction of hopping is reversed.
Going back to Eq.~(\ref{eqn-qs}), we get:

{\small\begin{equation}
Q\cdot S\left(\begin{array}{c}{\bf 1}\\{\bf 1}\end{array}\right)=
\overline{Q}\left(\begin{array}{c}{\bf 1}\\{\bf 1}\end{array}\right)=
^{L-2}C_{N-1}\left(\begin{array}{c}{\bf 1}\\{\bf 0}\end{array}\right)=
Q\left(\begin{array}{c}{\bf 1}\\{\bf 1}\end{array}\right)
\end{equation}}

\noindent All the conditions in the above equation can be satisfied if
we take $S\cdot{\bf 1}={\bf 1}$

We now analyze Eq.~(\ref{eqn-sp}) to put further constraints
on $S$.  Since both ${\bf P}$ and ${\bf \overline{P}}$ are
normalized, we obtain ${\bf 1}\cdot{\bf P}={\bf 1}\cdot S
\cdot {\bf \overline{P}}=1$ This equation can be satisfied
if we take ${\bf 1}\cdot S={\bf 1}+{\bf A}$, with the
additional constraint that ${\bf A}\cdot{\bf
\overline{P}}=0$. We observe that the equation is satisfied
only up to a vector ${\bf A}$ which is `orthogonal' to ${\bf
\overline{P}}$.

Acting by $W$ on both sides of Eq.~(\ref{eqn-sp}), gives:

\begin{equation} 
W\cdot{\bf P}=WS\cdot {\bf \overline{P}}=0
\end{equation}

\noindent This equation can be satisfied if we take
$WS=\pm\overline{W}$. The sign can be fixed by observing
that $WS\cdot {\bf 1} = W\cdot{\bf
1}=\pm\overline{W}\cdot{\bf 1}$. Where we have used the
property that $S\cdot{\bf 1}={\bf 1}$. Since the direction
of hopping has been reversed in ${\bf \overline{R}}$, the
outgoing configurations {\it w.r.t.}  a given initial
particle configuration are mapped to the incoming
configurations in ${\bf R}$ and vice versa, we obtain
$W\cdot{\bf 1}= - \overline{W}\cdot {\bf 1}$ (refer to
Eqs.~(\ref{eqn-transition} $\&$ \ref{eqn-rate})).

Thus the two equations that $S$ should satisfy are:
\begin{mathletters}
\label{eqn-sym}
\begin{eqnarray}
QS&=&\overline{Q} 
\label{eqn-symqs}\\
WS&=&-\overline{W}
\label{eqn-symws}
\end{eqnarray}
\end{mathletters}
\noindent An important feature to be noted is that it is not
a similarity transformation that relates the two systems
defined by $W$ and $\overline{W}$. By adding the two
equations, we obtain:

\begin{equation}
S=(Q+W)^{-1}\cdot(\overline{Q}-\overline{W})
\label{eqn-sols}
\end{equation}

Choosing $\eta=-1$ in Eq.~(\ref{eqn-eta}) and ${\bf
\overline{P}}=J_0 (\overline{Q}-\overline{W})^{-1}\cdot {\bf
V}$ with $1/J_0={\bf 1}\cdot
(\overline{Q}-\overline{W})^{-1}\cdot {\bf V}$ from
Eq.~(\ref{eqn-sol}), it can be shown that $S$ indeed
satisfies Eq.~(\ref{eqn-qp}).  This completes our proof that
the steady-state currents for ${\bf R}$ and ${\bf
\overline{R}}$ are the same. 

In conclusion, following an algebraic procedure, we have
derived formally exact expressions for the steady-state
probability density ${\bf P}$ and the current $J_0$ for the
totally asymmetric disordered simple exclusion process in
one dimension. We explicitly show how these solutions can be
exploited to study exact symmetries that leave the
steady-state current invariant. In particular we have
demonstrated that the steady-state current, up to a sign, is
the same when the direction of allowed jumps is reversed. We
also explicitly show that in steady-state the current across
each bond is the same constant, a result intuitively obvious
but lacking proof for disordered lattice models.  Although
not showed explicitly here, the choice of $S$ in
Eq.~(\ref{eqn-qp}) is a special case of the more general
requirement that $Q\cdot {\bf P}={\bf\overline Q}\cdot
{\bf\overline P}$, which guarantees the currents to be the
same under a symmetry operation. Hence, we believe that the
methodology we have developed has  wider applicablility in
studying and understanding the symmetries of more general
models not amenable to direct analytical studies.  

On completion of this work, we were made aware of an
alternate proof of the time reversal symmetry of the DTASEP
model given in Ref.~\cite{ref:goldstein/speer:pc} which in
contrast to our algebraic proof is probabilistic in
character. 

We would like to thank Mustansir Barma  and Goutam Tripathy
for bringing this problem to our notice and a careful
reading of the manuscript. We thank Drs.~S.~Goldstein and E.
Speer for sharing their unpublished results with us.  One of
us (AP) acknowledges very fruitful discussions with Chandan
Das Gupta, Toby Joseph and Amit Puniyani and KMK is greatful
to the Department of Science and Technology, India,
(DST/PAM/GR/381) for financial assistance.

\end{multicols} 
\end{document}